\documentclass[runningheads]{llncs}
\usepackage[T1]{fontenc}
\usepackage{graphicx}
\usepackage{listings}
\usepackage[ruled,vlined,linesnumbered]{algorithm2e}
\usepackage{amsmath}
\usepackage{hyperref}
\usepackage{xcolor}
\usepackage{caption}
\usepackage{float}

\SetCommentSty{mycommfont}
\begin{document}
\title{Pairwise and Attribute-Aware Decision Tree-Based Preference Elicitation for Cold-Start Recommendation}
\titlerunning{Pairwise Decision Tree-Based Preference Elicitation for recommendation}
%
\author{Alireza Gharahighehi\inst{1,2}
\and
Felipe Kenji Nakano\inst{1,2}\and
Xuehua Yang\inst{3} \and Wenhan Cu\inst{3} \and Celine Vens \inst{1,2}}
\authorrunning{A. Gharahighehi et al.}
%
\institute{KU Leuven, Campus Kulak, Department of Public Health and Primary Care, Kortrijk, Belgium \and
Itec, imec research group at KU Leuven, Kortrijk, Belgium
\\
\and
KU Leuven, Department of Mathematics, Leuven, Belgium\\
}
%






\maketitle              
\begin{abstract}
Recommender systems (RSs) are intelligent filtering methods that suggest items to users based on their inferred preferences, derived from their interaction history on the platform. Collaborative filtering-based RSs rely on users' past interactions to generate recommendations. However, when a user is new to the platform—referred to as a cold-start user—there is no historical data available, making it difficult to provide personalized recommendations. To address this, rating elicitation techniques can be used to gather initial ratings or preferences on selected items, helping to build an early understanding of the user's tastes. Rating elicitation approaches are generally categorized into two types: non-personalized and personalized. Decision tree-based rating elicitation is a personalized method that queries users about their preferences at each node of the tree until sufficient information is gathered. In this paper, we propose an extension to the decision tree approach for rating elicitation in the context of music recommendation. Our method: (i) elicits not only item ratings but also preferences on attributes such as genres to better cluster users, and (ii) uses item pairs instead of single items at each node to more effectively learn user preferences. Experimental results demonstrate that both proposed enhancements lead to improved performance, particularly with a reduced number of queries.

\end{abstract}
\section{Introduction}

Recommender Systems (RSs) are intelligent filtering systems that suggest items based on users' preferences. These systems utilize user logs and explicit feedback—such as ratings and likes/dislikes—to model individuals' preferences. There are two main categories of RSs: Content-Based Filtering (CB) and Collaborative Filtering recommender systems (CF)~\cite{lu2015recommender}. CB systems recommend items whose features match a user's profile, while CF systems leverage user logs to identify similarities between users and items in order to generate recommendations. Although CF systems generally outperform CB systems~\cite{adomavicius2005toward}, they suffer from the cold-start problem~\cite{gharahighehi2022addressing}. This issue arises when a new user joins the system and due to the lack of historical data, CF systems cannot reliably model the user's preferences.

One way to address the cold-start problem is through Rating (preference) Elicitation (RE). The idea is to explicitly ask new users to provide their preferences on a list of items. Based on these elicited preferences, CF can generate more reliable recommendations. RE algorithms rank items to gather user preferences, prioritizing those that are most informative for the CF. The more informative an item's rating is, the earlier it will be shown to new users during the RE process. This concept aligns with active learning~\cite{prince2004does} in machine learning, wherein an expert (the user, in this context) is queried to label (rate) the most informative instances (items)~\cite{elahi2016survey}. 

RE approaches are generally categorized into non-personalized and personalized methods. Non-personalized approaches—such as popularity-~\cite{rashid2002getting}, variance-, entropy-based~\cite{merialdo2001improving}, and HELF\footnote{Harmonic mean of Entropy and Logarithm of Frequency}~\cite{rashid2008learning}, elicit ratings from new users using the same list of items, ranked according to different heuristics. Conversely, personalized RE methods, such as IGCN\footnote{Information Gain through Clustered Neighbors}~\cite{rashid2008learning} and decision tree-based approaches~\cite{golbandi2011adaptive,zhou2011functional}, generate individualized item lists for each user. Some hybrid approaches combine non-personalized and personalized strategies, as seen in~\cite{gharahighehi2022adaptive,rubens2007influence}. Decision tree-based~\cite{golbandi2011adaptive} method have demonstrated better performance compared to non-personalized alternatives. It is a personalized RE algorithm that employs a ternary decision tree to select items for elicitation and ultimately cluster users based on their expressed preferences at the leaf nodes. In this type of decision tree, each user is presented with an item, and based on their response—like, dislike, or unknown—they traverse the tree to the next item.

In this paper, we aim to address the following two research questions regarding the decision tree-based RE algorithm: (i) Would it be more effective to elicit not only item ratings but also preferences on attributes—such as genre preferences—to cluster users more efficiently?
(ii) What would be the impact of using pairs of items, instead of single items, per node in the decision tree-based RE algorithm?

The contribution of this paper is two-fold. First, we demonstrate that eliciting user preferences not only on the final recommended item but also on item attributes during the initial onboarding of cold-start users is more effective for CF methods. Second, we show that eliciting user preferences over pairs of items, rather than single items, provides more informative signals for CF recommenders.

The paper is organized as follows: In Section~\ref{se:method}, the background and the proposed RE approaches are presented. Section~\ref{se:es} describes the music dataset used and the experimental setup for designing and evaluating the proposed approach. In Section~\ref{se:Re}, we present and discuss the preliminary results of applying our proposed approaches to the music dataset. Finally, Section~\ref{sec:conclusion} concludes the paper and outlines directions for future work.

\section{Methodology}
\label{se:method}
In this section\footnote{The source code is publicly shared in \url{https://github.com/camilleecu/master_thesis/tree/pairwise}}, we illustrate the simulation of RE, explain the mechanism of decision tree-based RE, and discuss its adaptation to elicit preference based on item attributes and item pairs.

\subsection{Rating Elicitation}
\begin{figure*}[h]
  \centering
  \includegraphics[width=\linewidth]{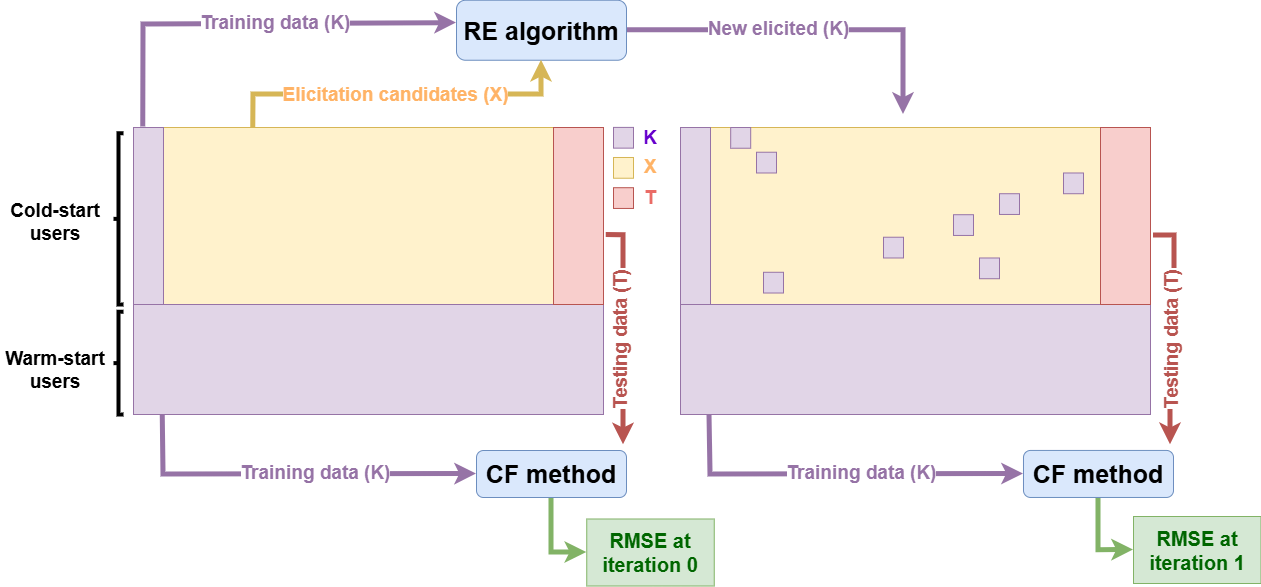}
  \caption{Illustration of RE procedure}\label{fig:data}
\end{figure*}

In order to assess the merits of different active learning methods for RE, we need to simulate the RE procedure based on users' historical logs. We follow the algorithm proposed by~\cite{elahi2014active} and partially adapt it to better imitate the real online experience, as described in Algorithm~\ref{alg:alg} and Figure~\ref{fig:data}. The RE algorithm takes as input the dataset containing users' ratings on items $\mathbf{D}$, a fixed number of elicitation iterations $\mathbf{N}$, and a base CF $\mathbf{R}$. Unlike the original RE algorithm proposed in~\cite{elahi2014active}, which initially considers all users as cold-start users, we divide users' logs into warm-start $\mathbf{D_{W}}$ and cold-start $\mathbf{D_{C}}$ and apply the RE steps to the cold-start dataset $\mathbf{D_{C}}$.  $\mathbf{D_{C}}$ is split into three disjoint datasets: $\mathbf{K}$, $\mathbf{X}$, and $\mathbf{T}$. $\mathbf{K}$ consists of a few ratings provided by cold-start users and, together with $\mathbf{D_{W}}$, is used to train the base RS $\mathbf{R}$. In other words, these are the ratings accessible to the CF. Ratings in $\mathbf{T}$ serve as test data and are used to evaluate the performance of the CF. The remaining ratings, $\mathbf{X}$, are used to apply active learning strategies for RE. If the item selected for RE by the active learning strategy exists in $\mathbf{X}$, it will be added to $\mathbf{K}$ in the next iteration. The more effective the active learning strategy for RE is, the more ratings from $\mathbf{X}$ will be transferred to $\mathbf{K}$, thereby improving the performance of $\mathbf{R}$.

\begin{algorithm}
\caption{Rating Elicitation for Collaborative Filtering Recommender Systems}
\label{alg:alg}

\KwIn{dataset $\mathbf{D}$, number of iterations $\mathbf{N}$, base recommender $\mathbf{R}$}
\KwOut{error per iteration $\mathbf{E}$}

\BlankLine
$D_{C}, D_{W} \gets U\_split(D)$ \tcp*[l]{Split users' logs based on cold- and warm-start users}  
$K, X, T \gets RE\_split(D_{C})$ \tcp*[l]{$K$ for training, $X$ for rating elicitation and $T$ for evaluation.}
$K \gets K \cup D_{W}$ \tcp*[l]{Add $D_{W}$ to the logs that system knows $K$}
$R \gets R.fit(K)$ \tcp*[l]{Fit $R$ only with $K$}
$E \gets R.evaluate(T)$\\
$j \gets 0$\\
   \While{$j<= N$}{
   $ A \gets$ $DT(K)$ \tcp*[l]{See Section~\ref{Me:dt}}
   $ K \gets K \cup (A\cap X)$ \tcp*[l]{The elicited ratings $A$ will be added to the known ratings $K$}
   $ X \gets X / A$ \tcp*[l]{The elicited ratings $A$ will be removed from rating elicitation candidates $X$}
   $R \gets R.fit(K)$ \tcp*[l]{Fit $R$ with the enhanced training data $K$}
   $E \gets R.evaluate(T)$\\
   $j \gets j +1$ \\
}
\textbf{Return} $E$
\end{algorithm}

\subsection{Decision Tree based rating elicitation}
\label{Me:dt}
The decision tree method for RE~\cite{golbandi2011adaptive} is a ternary decision tree. At each node, users are presented with an item, and based on their ratings, they are clustered into three groups: lovers, who give high ratings to the item; haters, who give low ratings; and unknowns, who do not rate the item. If further elicitation steps are needed, each of the three resulting branches presents another item to the corresponding group of users. To train the decision tree—i.e., to determine which items to use for splitting users—each candidate item is evaluated by calculating the prediction error if it were used at a node. The item with the lowest prediction error ($E(j)$), calculated as follows, is selected for the split:

\begin{equation}
E(j)=e^{2}(tL)+e^{2}(tH)+e^{2}(tU)
\label{eq:E}
\end{equation}
\begin{equation}
e^{2}(t)=\sum_{i\in I}\sum_{u \in U_{t}\cap U_{i}} (r_{ui}-\mu(t)_{i})
\label{eq:e}
\end{equation}

\noindent where $e^{2}(tL)$, $e^{2}(tH)$, and $e^{2}(tU)$ are squared error for the lovers, haters and unknowns nodes respectively, if tree is split based on item $j$. In Equation.~\ref{eq:e}, $I$ is the set of all items, $U_{t}$ is the set of all users in node t, $U_{i}$ is the set of users who have rated item $i$, $r_{ui}$ is the real rating that user $u$ gave to item $i$, and $\mu(t)_{i}$ is the mean ratings that users in node $t$ gave to item $i$. $\mu(t)_{i}$ can also be seen as the predicted rating of item $i$ by users who are located in node $t$. 
When the item with lowest error for split is found, three branches will be made and for each branch, the same procedure will be applied to find the next item to split users. Since the source code for the original decision tree-based rating elicitation (RE) method proposed by Golbandi et al.~\cite{golbandi2011adaptive} is not publicly available, and the paper implies that the decision tree trained on the initial users' logs $K$ (In our setting, this includes the initial logs of cold-start users as well as the logs of warm-start users as indicated in line 3 of Algorithm~\ref{alg:alg}) remains fixed throughout all iterations, we have developed a modified version of the RE algorithm that incorporates two key enhancements:

\begin{itemize}
    \item \textbf{Dynamic tree updating:} In each RE iteration (Line 8-12 in Algorithm~\ref{alg:alg}), the decision tree is retrained using the updated known ratings $K$ (line 9 in Algorithm~\ref{alg:alg}), allowing the model to adapt to newly acquired user ratings in previous RE iterations.
    \item \textbf{Top-Down Rating Elicitation:} Rather than selecting an item at a depth corresponding to the iteration number, we initiate the elicitation process from the root of the updated decision tree in each iteration. We then traverse the tree top-down, selecting the first item whose ratings have not yet been elicited from the user. 
\end{itemize}

These modifications aim to improve the adaptability and efficiency of the RE process by leveraging updated user feedback in each iteration.







\subsubsection{Decision Tree based rating elicitation with multiple item types}

In some application domains, users can express preferences not only on items, but also on item attributes. For example, on certain music platforms, users may rate not only music tracks but also artists, genres, and albums. In such cases, a candidate item at each node of the decision tree can belong to any of these categories. While the final recommendation task may focus on a specific category, such as music tracks, eliciting ratings from other categories can provide more informative signals for the CF. Therefore, items from these alternative categories may be prioritized during the rating elicitation process. These items are evaluated using Equations (1) and (2) and may be selected at decision tree nodes instead of items from the primary category.

\subsubsection{Decision Tree based rating elicitation with item pairs}

Previous research has shown that humans are generally better at making comparative judgments between pairs of items than at evaluating individual items in isolation~\cite{lesterhuis2017comparative}. Based on this insight, the decision tree for rating elicitation (RE) can be constructed using item pairs instead of single items at each node. In this approach, each step of the RE process presents users with a pair of items. Depending on their response, users proceed to one of three branches in the decision tree: those who prefer the first item, those who prefer the second, and those who are indifferent between the two. Since the number of possible item pairs grows rapidly with the number of items, we introduce two heuristics to efficiently select item pairs at each node. First, we identify the list of top $K$ items with the lowest prediction error $E(j)$, and then form pairs among the same item types using one of the following strategies:

\begin{itemize}
    \item Pair the first two items in the list (\texttt{pairwise\_tree\_1});
    \item Pair the first item with the most similar item in the top k list (\texttt{pairwise\_tree\_2});
\end{itemize}

Item similarity is computed using the rating matrix, with cosine similarity as the similarity measure. The rationale behind the first strategy is that the two items with the lowest prediction error are considered the most informative, and therefore, their combination is likely to be highly informative. The second strategy assumes that the first item is already highly informative, and aims to find another item that is also informative (i.e., among the top $K$ items with the lowest prediction error), but more likely to be comparable with the first. Since similarity is calculated based on the overlap in users who rated each item, a similar item is one that has been rated by users with similar preferences. This increases the likelihood that the user can meaningfully compare the items in the pair.

\section{Experimental Setup}
\label{se:es}
In this paper we used Yahoo! music dataset to evaluate our proposed RE approaches. To the best of our knowledge, this is the only music dataset that contains user ratings on different item types, such as music tracks, artists, genres, and albums. The whole dataset contains around one million users, 600k unique items (including all four types) and 250 million ratings, all between 0 and 100. The main recommendation task we consider is artist recommendation, since typical CF methods are not suitable for music track recommendation, due to the sequential nature of track recommendation. To answer to our first research question, we consider genre as the other item type that can be used in RE. In order to get dense enough initial matrix for our experiments, we drop users with less than 100 interactions, artists with less than 500 and genres with less than 500 ratings. One might argue that users with at least 100 ratings should not be considered cold-start. To clarify, this threshold is used solely to construct the initial dataset. In subsequent steps, we extract a small subset of this data to form $K$, which is used, together with the warm users' logs $D_{W}$, to fit the CF, denoted as R in Algorithm~\ref{alg:alg}. The remaining data is used to simulate rating elicitation for cold-start users, those who have very few ratings in $K$, and to evaluate the effectiveness of our active learning strategy. The final dataset that we used contains 1,701 users, 4,793 artists and 218 genres.

As mentioned in Algorithm~\ref{alg:alg} (Line 1), to better mimic the real situation, we split the users into cold-start and warm users. Since our main focus in this paper is cold-start users, in the experiments we consider this ratio as 90\%. To form $K$, $X$, and $T$ datasets (Line 2 in Algorithm~\ref{alg:alg}), we considered one artist rating for each user in the $K$, 30 ratings for $T$, and all the rest ratings for $X$. The number of RE iterations is fixed to 20 in the experiments. In this paper, we use Singular Value Decomposition (SVD) as the base CF. Naturally, other CF could also have been applied, nonetheless our focus relies on investigating the effect of RE strategies on the performance of the CF method. To evaluate the performance of RE approaches, we use Root Mean Squared Error (RMSE) measure, since it is vastly employed in the literature~\cite{elahi2016survey}.


To apply our proposed pairwise decision tree approach, the dataset must first be pre-processed. Since the data was originally collected as ratings (scaled from 0 to 100) on individual items, we transform the rating matrix into a semi-binary format to better reflect a pairwise RE scenario. Specifically, if a user's rating is equal to or greater than 50, we set the value to 1; if the rating is below 50, we assign a small value (0.01); otherwise, the rating is treated as missing. We assigned 0.01 to avoid confusing it with missing ratings, as some RSs, such as SVD, treat the missing values as zero. Based on these transformed values, users are clustered into three branches according to their preferences for item pairs at each decision tree node. If both items have the same transformed rating—either 1, 0.01, or missing—the user is directed to the indifference branch. If the first item is rated 1 and the other is either missing or rated 0.01, the first item is considered preferred. Conversely, if the second item is rated 1 and the first is missing or rated 0.01, the second item is preferred. Finally, in pairs with missing and 0.01 rating, the latter will be preferred, since rated items are generally assumed to be preferred over unrated ones~\cite{rendle2012bpr}. To construct pairs, top 20 items with the lowest prediction error in each node of the decision tree are considered.

\section{Results and Discussion}
\label{se:Re}

\begin{figure}
    \centering
    \includegraphics[width=\linewidth]{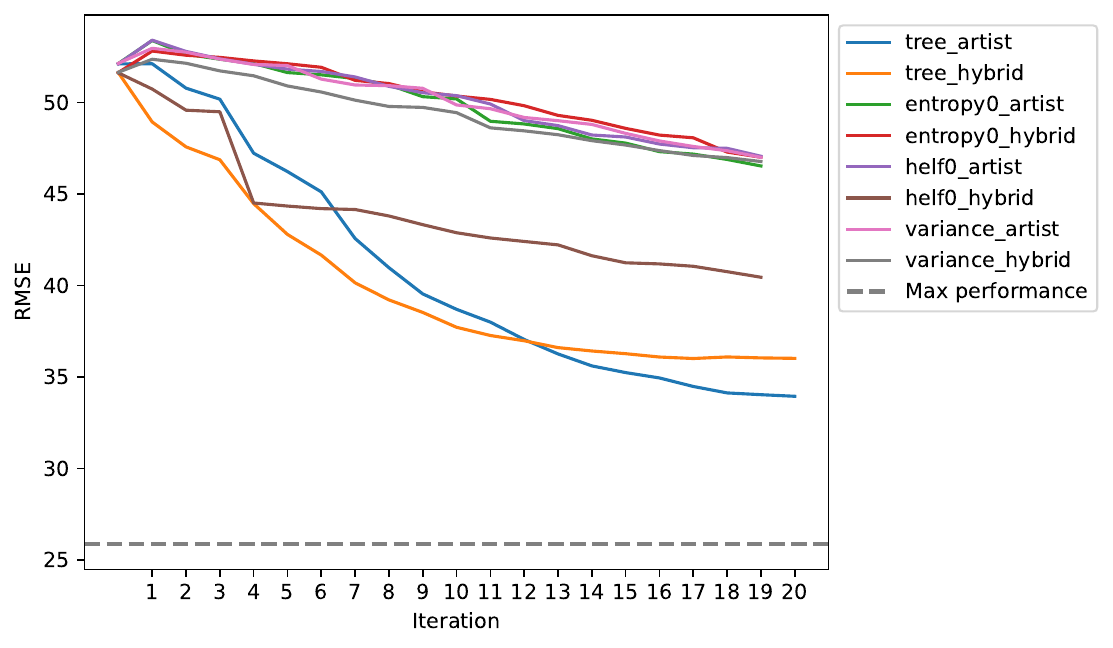}
    \caption{Decision tree rating elicitation with multiple categories of items}
    \label{re:RQ1}
\end{figure}

\begin{figure}
    \centering
    \includegraphics[width=\linewidth]{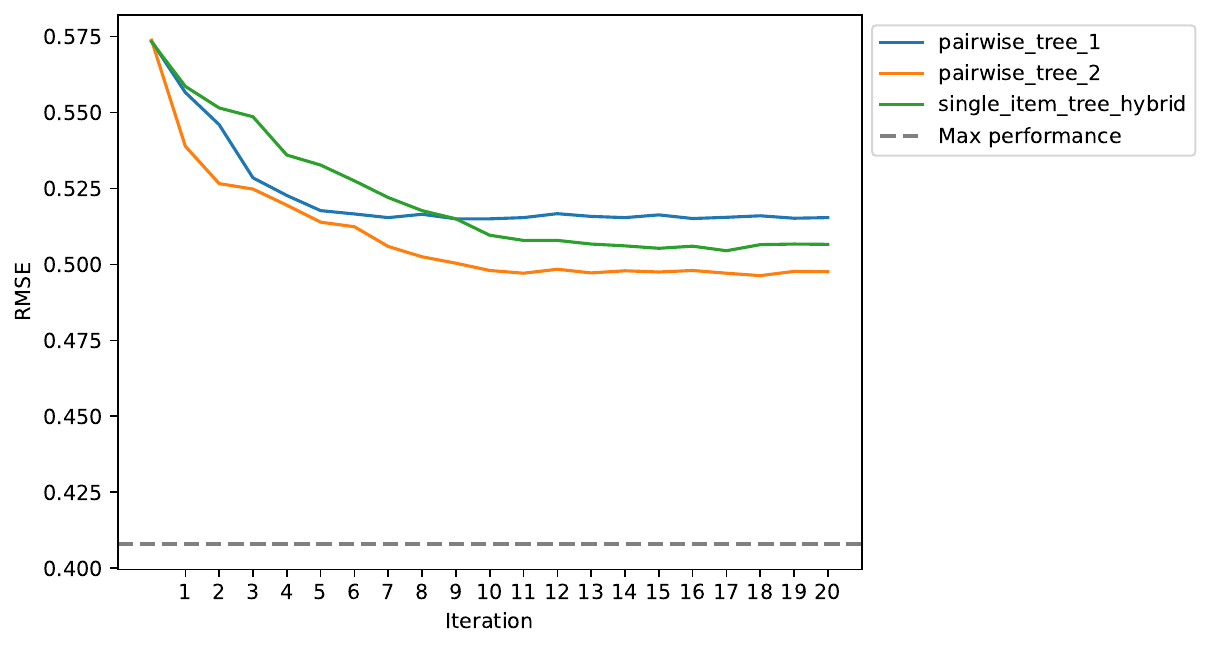}
    \caption{Decision tree rating elicitation with pairs of items}
    \label{re:RQ2}
\end{figure}

The preliminary results of applying the proposed decision tree-based RE approaches are reported in Figure~\ref{re:RQ1} and~\ref{re:RQ2}. As shown in Figure~\ref{re:RQ1}, the decision tree-based approach that elicits users' preferences on both artists and genre ($tree\_hybrid$) provides more informative input to an artist RS compared to the best non-personalized approach, $helf0\_hybrid$. Starting from iteration 13, the decision tree-based model that elicits preferences only on artists outperforms the hybrid approach. This implies that once we have sufficient knowledge about the users, it is more effective to elicit preferences on the target item type rather than on item attributes. However, since cold-start users are typically reluctant to answer too many questions during RE procedure, the proposed $tree\_hybrid$ approach appears to perform best for such users.


Figure~\ref{re:RQ2} illustrates the results related to our second research question. The RMSE scale in Figure~\ref{re:RQ1} is much larger than in this figure, as the rating scale is also much higher (0–100 compared to 0–1). As shown in Figure~\ref{re:RQ2}, the pairwise decision tree-based RE approach ($pairwise\_tree\_2$) outperforms the decision tree-based RE approach with a single item per node ($single\_item\_tree\_hybrid$). This suggests that eliciting preferences on pairs of items provides more informative signals to the CF, thereby requiring fewer RE requests from users. Consistent with the literature on comparative judgment, this result also implies that users provide more informative feedback when comparing item pairs than when evaluating individual items.

While our results successfully addressed our research questions, we acknowledge some limitations in our work. First, we relied on a single dataset to conduct our analysis. To the best of our knowledge, the music dataset we used is the only one available that includes user ratings across multiple item types. Another limitation is that our proposed approaches were evaluated solely using historical user logs. Due to the lack of access to end users, we were unable to conduct an online evaluation approach.


\section{Conclusion}
\label{sec:conclusion}
In this paper, we present results from our research on decision tree-based rating elicitation for music. Our contribution is two-fold:
(i) We show that a decision tree-based rating elicitation algorithm that gathers both item and attribute ratings outperforms one that collects only item ratings, particularly during early user interactions. (ii) We show that a decision tree constructed using pairs of items outperforms one based on individual items with the same number of interactions with the users.

There are three main directions for future work. While our preliminary results suggest that the pairwise approach is more effective for preference elicitation compared to single-item rating elicitation, we relied on simple heuristics to compose item pairs. Further research is needed to find the optimal item pair at each node of the decision tree. In the proposed pairwise decision tree approach, the base collaborative filtering recommender system is trained using inferred semi-binary ratings. An alternative would be to use pairwise preferences to train recommendation systems with learning-to-rank loss functions, such as Bayesian Personalized Ranking (BPR)~\cite{rendle2012bpr}. This would allow the elicited preferences to be directly incorporated into the training of the recommender model. In this study, users' historical logs were used to simulate the rating elicitation scenario. However, if access to end users is available, the proposed approaches could also be evaluated in an online setting. Finally, instead of a single RE method, an ensemble of recommenders~\cite{gharahighehi2021ensemble} could be applied to provide more robust elicitation.

\section*{Acknowledgment}
The authors acknowledge support from the Flemish Government (AI Research Program) and Research Fund Flanders (FWO) mandate 1235924N.  

\bibliographystyle{splncs04}
\bibliography{references}

\end{document}